\documentclass[aps,prb,amsmath,amssymb,twocolumn,superscriptaddress]{revtex4-1}

\usepackage{graphicx}
\usepackage{dcolumn}
\usepackage{bm}

\setcitestyle{numbers,square}

\usepackage{graphicx}
\usepackage{dcolumn}
\usepackage{bm}
\usepackage{hyperref}

\begin{document}

\title{Magnetism in spin crossover systems: short-range order and effects beyond the Heisenberg model}

\author{V. I. Kuz'min}
\email{kuz@iph.krasn.ru}
\affiliation{Kirensky Institute of Physics, Federal Research Center KSC SB RAS, 660036 Krasnoyarsk, Russia}
 
\author{Yu. S. Orlov}
 \affiliation{Kirensky Institute of Physics, Federal Research Center KSC SB RAS, 660036 Krasnoyarsk, Russia}
 \affiliation{Institute of Engineering Physics and Radio Electronics, Siberian Federal University, 660041 Krasnoyarsk, Russia}

\author{A. E. Zarubin}
 \affiliation{Institute of Engineering Physics and Radio Electronics, Siberian Federal University, 660041 Krasnoyarsk, Russia}
\author{T. M. Ovchinnikova}
\affiliation{Sukhachov Institute of Forest, Federal Research Center KSC SB RAS, 660036 Krasnoyarsk, Russia}
\author{S. G. Ovchinnikov}
 \affiliation{Kirensky Institute of Physics, Federal Research Center KSC SB RAS, 660036 Krasnoyarsk, Russia}
 \affiliation{Institute of Engineering Physics and Radio Electronics, Siberian Federal University, 660041 Krasnoyarsk, Russia}

\date{\today}

\begin{abstract}
To study non-Heisenberg effects in the vicinity of spin crossover in strongly correlated electron systems we derive an effective low-energy Hamiltonian for the two-band Kanamori model. It contains Heisenberg high-spin term proportional to exchange constant as well as low-spin term proportional to spin gap parameter $\varepsilon_s$. Using cluster mean field theory we obtain several non-Heisenberg effects. Near critical value of spin gap $\varepsilon^c_s$ there is a magnetic phase transition of first order. In the vicinity of $\varepsilon^c_s$ in the paramagnetic phase we observe non trivial behavior of the Curie constant in the paramagnetic susceptibility in the wide range of temperature. Reentrant temperature behavior of nearest-neighbor spin-spin correlations is observed at $\varepsilon_s > \varepsilon^c_s$. Finally, pressure-temperature magnetic phase diagram for ferroperriclase is obtained using the effective Hamiltonian.
\end{abstract}

\maketitle

\section{Introduction}

Spin crossover (SCO) is a phenomenon which takes place when the metal ion changes its spin state between low spin (LS) and high spin (HS) configuration under the effect of external perturbation such as pressure, magnetic field, temperature, or light irradiation. The SCO can be observed in transition metal compounds (often in the 3d-metal oxides with $d^4$-$d^7$ electronic configurations) \cite{Halder, Brooker, LyubutinUFN, Ohkoshi} or in transition metal complexes, like metalorganic molecules or molecular assemblies \cite{Saha-Dasgupta}. Free inertial molecular switches to store and process information in fast computational devices were the primary interest for SCO. In the nanotechnology certain properties of the SCO are of the interest for quantum transport and a new generation of sensors and displays \cite{Jureschi}. The SCO in Fe-containing oxides is also important for the understanding the physical properties of the Earth’s mantle \cite{Wentzcovitch, Hsu, Liu, Ovchinnikov12, Sinmyo}.

At first glance the SCO is a problem of an individual ion and results from the competition of the Hund intra-atomic exchange interaction and the crystal field value determined by surrounding ions. Nevertheless, the effective interaction between magnetic ions due to electron-phonon, exchange, and quadrupole couplings results in cooperative effects, which provide different hysteresis phenomena and play an important role in practical applications and understanding the origin of the SCO. There are many papers where the cooperative effects have been treated within the Ising model \cite{Jureschi, Wajnflasz, Bari, Nishino03, Timm, Banerjee, Paez-Espejo}. In all these studies the effective exchange interaction is postulated phenomenologically within the Ising or Heisenberg model with empirical exchange parameters. In the last decade the cooperative effects in SCO have been studied by the density functional theory \cite{Marbeuf}, molecular dynamics \cite{Nishino07, Boukheddaden}, and Monte Carlo simulations \cite{Konishi, Miyashita}. The interplay of electron hopping between neighboring ions with the orbital structure of different spin multiplets also results in spin-orbital cooperative effects in strongly correlated transition metal oxides \cite{Sboychakov}.

In conventional magnetic insulators only the ground term $E_0$ of magnetic cation in the multielectron configuration $d^n$ with some spin value $S_0$ is involved in the formation of the Heisenberg Hamiltonian as the effective low-energy model. The important difference of the magnetism in SCO systems is that at least two different terms, usually HS and LS, are involved in the formation of the effective low energy model. This is a reason for the non-Heisenberg model effects that will be discussed in this paper. Recently we have developed a general approach to construct the effective exchange interaction model that takes into account the contribution of the excited terms of the magnetic cation \cite{Gavrichkov17} and found that the interatomic exchange interaction results in the SCO to be the first order phase transition \cite{Nesterov}. For arbitrary $d^n$ configuration we cannot write down analytically the parameters of the effective Hamiltonian that contains the interatomic exchange as well as the interatomic hopping of excitons, the excitations between HS and LS terms.

In this paper we study more simple toy model with two electronic orbitals and the Coulomb interaction in the Kanamori approach \cite{Kanamori}. Within the generalized tight binding (GTB) method \cite{Ovchinnikov89, Gavrichkov00} to the electronic structure of strongly correlated systems we provide the exact diagonalization of the local intraatomic part of the Hamiltonian, construct the Hubbard operators using a set of the exact local eigenstates, and write down the total Hamiltonian as the multiorbital Hubbard model. This model describes a magnetic insulator with the energy gap $E_g$ between the occupied valence and empty conductivity bands. Two electrons per site form the HS triplet and LS singlets with the SCO at increasing the crystal field splitting between two orbitals (for example, by external pressure). We should mention that similar models under different names (the two-band Hubbard model or the extended Falicov-Kimball model) have been intensively discussed in the literature, see the review paper \onlinecite{Kunes15}. We write down explicitly the matrix elements of the exchange and exciton hopping contributions, which are beyond the conventional Heisenberg model. The other non-Heisenberg model effect is related to a structure of the local Hilbert space, which contains for our model 3 magnetic eigenstates for HS with $S=1$ and 3 singlets with $S=0$. Within the two-band Hubbard model similar strong coupling approach \cite{Werner, Suzuki, Nasu} has also revealed two terms with $S=1$ and $S=0$ for electronic concentration $n_e=2$ and the intersite interaction matrix elements (see also Refs.~\onlinecite{Balents, Kaneko, Kunes14}). The main object of all these papers is the possible excitonic condensation in systems of strongly correlated electrons. In our paper we restrict our interest to the SCO systems and possible non Heisenberg effects. The presence of the additional LS states does not allow introducing the Brillouin function in the mean field (MF) approximation. A small number of electrons in our toy model ($n_e=2$ per site) allows us to study the model’s phase diagram applying a cluster mean field (CMF) approach in order to go beyond the standard MF. In this way we can obtain qualitative information about the model's phase diagram and explore the validity of approximation by considering different cluster sizes as well as discuss the short-order effects, which are also different from the conventional Heisenberg model, in the vicinity of the first order transition from the HS antiferromagnetic phase into the LS non magnetic phase due to local nature of SCO.

The paper is organized as follows. In section \ref{sec:2} we describe the two-orbital Kanamori model, the effective low energy Hamiltonian containing HS and LS states, and interatomic exchange interaction and exciton hopping. In section \ref{sec:3} we briefly remind the CMF theory. The non-Heisenberg model and short-order effects in the vicinity of spin crossover are discussed in section \ref{sec:4}. In section \ref{sec:5} we discuss the main results. 

\section{\label{sec:2}Two-band Kanamori model}

The multielectron states for the $d^n$-configuration in the cubic crystal field can be obtained from the Tanabe-Sugano diagrams \cite{Tanabe, Sugano}, which demonstrate stability of the HS terms for small value of the crystal field $10Dq$, and that the crossover of the HS and LS terms takes place for $d^4$-$d^7$ electronic configurations with increasing the crystal field value stabilizing the LS state. Beyond the crystal field theory, the SCO may also happen due to increasing the cation-anion $p$-$d$ hybridization \cite{Ovchinnikov07}. The minimal multielectron model to discuss SCO is the two-orbital tight-binding model that includes two single electron levels $\varepsilon_1$ and $\varepsilon_2$ with interatomic hopping $t_{i,j}$ and the local Coulomb interaction for electron concentration $n_e=2$. Its Hamiltonian is given by
\begin {equation}
H = H_t + H_{Coulomb}.
\label {eq:1}
\end {equation}
The interatomic term
\begin {eqnarray}
H_t &=& \varepsilon_1 \sum\limits_{i, \sigma}{a_{i1\sigma}^{\dag}}{a_{i1\sigma}^{}} + \varepsilon_2 \sum\limits_{i, \sigma}{a_{i2\sigma}^{\dag}}{a_{i2\sigma}^{}} \nonumber \\
&+& t_1 \sum\limits_{\left\langle i, j \right\rangle, \sigma} {a_{i1\sigma}^{\dag}}{a_{j1\sigma}^{}} + t_2 \sum\limits_{\left\langle i, j \right\rangle, \sigma} {a_{i2\sigma}^{\dag}}{a_{j2\sigma}^{}} \nonumber \\
&+& t_{12} \sum\limits_{\left\langle i, j \right\rangle, \sigma}\left( {a_{i2\sigma}^{\dag}}{a_{j1\sigma}^{}} + {a_{i1\sigma}^{\dag}}{a_{j2\sigma}^{}} \right)
\label {eq:2}
\end {eqnarray}
describes the intraband $t_1$ and $t_2$ hoppings and the interband hopping $t_{12}$ of electrons between the nearest neighbor sites with the single electron energies $\varepsilon_1$ and $\varepsilon_2 = \varepsilon_1 + \Delta$, where $\Delta$ is the crystal field value. The local Coulomb interaction within the Kanamori approach contains different matrix elements, the intraorbital $U$ and interorbital $V$, as well as the Hund coupling $J$ and the interband coupling $J'$:
\begin{widetext}
\begin {eqnarray}
H_{Coulomb} = U \sum\limits_{i, \lambda}{a_{i\lambda\uparrow}^{\dag}}{a_{i\lambda\downarrow}^{\dag}}{a_{i\lambda\uparrow}^{}}{a_{i\lambda\downarrow}^{}} + V \sum\limits_{i, \lambda\neq\lambda'}{a_{i\lambda\uparrow}^{\dag}}{a_{i\lambda'\downarrow}^{\dag}}{a_{i\lambda\uparrow}^{}}{a_{i\lambda'\downarrow}^{}} + V \sum\limits_{i, \lambda>\lambda'}{a_{i\lambda\sigma}^{\dag}}{a_{i\lambda'\sigma}^{\dag}}{a_{i\lambda\sigma}^{}}{a_{i\lambda'\sigma}^{}} \nonumber \\
+ J \sum\limits_{i, \lambda>\lambda',\sigma}{a_{i\lambda\sigma}^{\dag}}{a_{i\lambda'\sigma}^{\dag}}{a_{i\lambda'\sigma}^{}}{a_{i\lambda\sigma}^{}} + J \sum\limits_{i, \lambda\neq\lambda'}{a_{i\lambda\uparrow}^{\dag}}{a_{i\lambda'\downarrow}^{\dag}}{a_{i\lambda'\uparrow}^{}}{a_{i\lambda\downarrow}^{}} + J' \sum\limits_{i, \lambda\neq\lambda'}{a_{i\lambda\uparrow}^{\dag}}{a_{i\lambda\downarrow}^{\dag}}{a_{i\lambda'\uparrow}^{}}{a_{i\lambda'\downarrow}^{}} 
\label {eq:3}
\end {eqnarray}
\end{widetext}
In the limit  $\Delta=0$ and for one electron per site this model transforms in the Kugel-Khomskii model for charge ordering \cite{Kugel82}. In this paper we will consider this model only for homopolar case $n_e=2$. As we have mentioned in the introduction, similar models have been studied recently to find the excitonic insulator phase.

\begin{figure}
\includegraphics{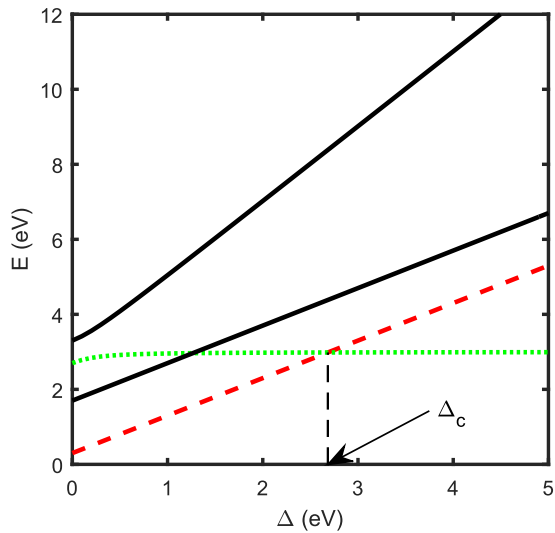}
\caption{\label{fig:1}The crystal field dependence of the two-electron local eigenstates. The red dashed line shows the ground HS term for $\Delta < \Delta_c$, the green dotted line indicates the ground LS term for $\Delta > \Delta_c$, black solid lines correspond to the high-energy singlets. Calculation has been carried out for the following parameters: $U=3 \text{eV}$, $V=1 \text{eV}$, $J=0.7 \text{eV}$, and $J=0.7 \text{eV}$.}
\end{figure}

For zero interatomic hopping there are 6 exact two-electron states. The triplet ($S=1$)
\begin{equation}
\left|\sigma\right\rangle=\begin{cases}
               a^{\dag}_{1\uparrow}a^{\dag}_{2\uparrow}\left|0\right\rangle, \sigma = +1 \\
               \frac{1}{\sqrt{2}}\left(a^{\dag}_{1\uparrow}a^{\dag}_{2\downarrow}+a^{\dag}_{1\downarrow}a^{\dag}_{2\uparrow}\right)\left|0\right\rangle, \sigma = 0 \\
               a^{\dag}_{1\downarrow}a^{\dag}_{2\downarrow}\left|0\right\rangle, \sigma = -1
            \end{cases}
\label {eq:4}
\end{equation}
triply degenerate HS-term $\left|\sigma\right\rangle$ with the energy $E_{HS}=2\varepsilon_1+\Delta+V-J$ is the ground state for the crystal field  (Fig.\ref{fig:1}, red dashed line), for $\Delta>\Delta_c$   the singlet ($S=0$) LS state 
\begin{equation}
\left|S\right\rangle=C_1\left(\Delta\right)a^{\dag}_{1\uparrow}a^{\dag}_{1\downarrow}\left|0\right\rangle - \sqrt{1-C^2_1\left(\Delta\right)}a^{\dag}_{2\uparrow}a^{\dag}_{2\downarrow}\left|0\right\rangle,
\label {eq:5}
\end{equation}
where $C_1\left(\Delta\right)=J'/\sqrt{{J'}^2 - \left(2\varepsilon_1+U-E_{LS}\right)^2}$, with the energy $E_{LS}=2\varepsilon_1+\left(\Delta+U\right)-\sqrt{\Delta^2-{J'}^2}$ becomes the ground state (Fig.\ref{fig:1}, green dotted line). The crossover occurs at $\Delta=\Delta_c=\sqrt{\left(U-V+J\right)^2-{J'}^2}$. There are two more singlets,
\begin{equation}
\left|S_1\right\rangle=\frac{1}{\sqrt{2}}\left(a^{\dag}_{1\uparrow}a^{\dag}_{2\downarrow}-a^{\dag}_{1\downarrow}a^{\dag}_{2\uparrow}\right)\left|0\right\rangle
\label {eq:6}
\end{equation}
with the energy $E_{S_1}=2\varepsilon_1+\Delta+V+J$ and 
\begin{equation}
\left|S_2\right\rangle=\left(\sqrt{1-C^2_1\left(\Delta\right)}a^{\dag}_{1\uparrow}a^{\dag}_{1\downarrow}+C_1\left(\Delta\right)a^{\dag}_{2\uparrow}a^{\dag}_{2\downarrow}\right)\left|0\right\rangle
\label {eq:7}
\end{equation}
with the energy $E_{S_2}=2\varepsilon_1+\left(\Delta+U\right)+\sqrt{\Delta^2-{J'}^2}$, which are excited for all parameters; they are shown by the solid black lines in Fig.\ref{fig:1}.

To treat the intersite electron hopping we use the GTB approach \cite{Ovchinnikov89, Gavrichkov00, Korshunov}, which is a version of cluster perturbation theory. We introduce the Hubbard $X$-operators $X^{pq}=\left|p\right\rangle\left\langle q\right|$, where where $\left|p\right\rangle$ and $\left|q\right\rangle$ are the eigenstates of the Hamiltonian (\ref{eq:1}) at $t_{\lambda\lambda'}=0$ with different numbers of electrons $n_e = 1,2,3$. A single electron creation/annihilation operator at site $i$ with an orbital index $\lambda$ as well as any other local operator is given by a linear combination of the Hubbard operators \cite{Hubbard}:
\begin{equation}
a_{i\lambda\sigma}=\sum\limits_{pq}\left|p\right\rangle\left\langle p\right\rangle a_{i\lambda\sigma} \left|q\right\rangle \left\langle q\right| = \sum\limits_{pq}\gamma_{\lambda\sigma}\left(pq\right)X^{pq}_i.
\label {eq:8}
\end{equation}
The number of different quasiparticles $\left(pq\right)$ is finite, one can numerate them by the number $m$, which is the quasiparticle band index, then $a_{i\lambda\sigma}=\sum\limits_m \gamma_{\lambda\sigma}\left(m\right)X^{m}_i$ $\left(a^{\dag}_{i\lambda\sigma}=\sum\limits_m \gamma^*_{\lambda\sigma}\left(m\right){X_i^{m}}^{\dag}\right)$.

In the $X$-operator representation the Hamiltonian (\ref{eq:1}) can be written exactly as 
\begin{equation}
H = \sum\limits_{i,p}E_p X^{pp}_i + \sum\limits_{\left\langle i,j\right\rangle}\sum\limits_{mn}t^{mn}{X_i^{m}}^{\dag} X^{n}_j.
\label {eq:9}
\end{equation}
Here $E_p$ is the energy of the term $\left|p\right\rangle$, $t^{mn} = \sum\limits_{\sigma,\lambda,\lambda'}t_{\lambda\lambda'}\gamma^*_{\lambda\sigma}\left(m\right)\gamma_{\lambda'\sigma}\left(n\right)$ is the intersite hopping matrix element. We would like to emphasize that the Hamiltonian (\ref{eq:9}) is the general multielectron Hamiltonian that is valid for any complete and orthonormalized set of local eigenstates, all microscopic details are given by the structure of local eigenstates.

For number of electrons $n_e=2$ the Hamiltonian (\ref{eq:9}) results in the Mott-Hubbard insulator ground state with the insulator band gap $E_g$. The localized magnetic moment at each site is HS for $\Delta<\Delta_c$ and LS for $\Delta>\Delta_c$. To obtain the interatomic exchange interaction we apply the method developed for the Hubbard model \cite{Chao} and generalized for arbitrary set of local eigenstates in \cite{Gavrichkov17} (see also Refs. \onlinecite{Kunes15, Nasu}). The idea is to construct the effective Hamiltonian excluding the interband interatomic hopping. Contrary to the general case, in our toy model we can write down the exchange interaction analytically. The effective Hamiltonian is equal to
\begin{equation}
H_{eff} = H_{s} + H_{ex}.
\label{eq:10}
\end{equation}
Here the first term is the spin Heisenberg-type Hamiltonian, while the second term describes the non-Heisenberg intersite hopping of the local excitons. This Hamiltonian acts within the Hilbert space that contains four states: three $S=1$ triplet states $\left|-\right\rangle$, $\left|0\right\rangle$, $\left|+\right\rangle$ and the singlet state $\left|s\right\rangle$. The spin part is given by
\begin{equation}
H_{s} = \frac{J}{2} \sum\limits_{\left\langle i,j\right\rangle}\left({\bf S}_i {\bf S}_j-\frac{1}{4}n_in_j\right)-\varepsilon_s \sum\limits_i X^{ss}_i,
\label{eq:11}
\end{equation}
where the superexchange parameter is 
\begin{equation}
J=4\left(t^2_{11}+2t^2_{12}+t^2_{22}\right)/E_g,
\label{eq:12}
\end{equation}
${\bf S}_i$ is the $S=1$ spin operator, in the Hubbard operators given by $S^+_i=\sqrt{2}\left(X^{+0}_i + X^{0-}_i\right)$, $S^-_i=\sqrt{2}\left(X^{0+}_i + X^{-0}_i\right)$, $S^z_i=\sqrt{2}\left(X^{++}_i - X^{--}_i\right)$, and $n_i=q_e\left(X^{++}_i + X^{--}_i + X^{00}_i + X^{ss}_i\right)$ is the number of electrons operator, $q_e=2$ is the number of electrons per site, in our homopolar case the completeness of our two-electron exact set of eigenvectors looks like
\begin{equation}
X^{++}_i + X^{--}_i + X^{00}_i + X^{ss}_i = 1,
\label{eq:13}
\end{equation}
so $n_i=2$. The last term in the Hamiltonian $H_s$ (\ref{eq:11}) is the non-Heisenberg contribution of the nonmagnetic LS state with the spin gap value $\varepsilon_s=E_{HS}-E_{LS}$. This is the local exciton energy. Below we will assume the linear dependence of the crystal field parameter on the external pressure: $\Delta = \Delta(0) + aP$  due to the linear decrease of crystal volume under the pressure.

The creation/annihilation of the local excitons is given by the Hubbard operators $X^{\sigma s}_i$ (from the initial LS state $\left|s\right\rangle$ in the final HS state $\left|\sigma\right\rangle$, and $X^{s\sigma}_i$ corresponds to the back excitation. These excitons describe the fluctuations of multiplicity, the term used many years ago in the paper \cite{Vonsovskii}. We consider this term is the appropriate one in the spin crossover physics, the term spin fluctuations in magnetism usually means the change of a spin projection for the same value of the spin. The second part of the effective Hamiltonian (\ref{eq:10}) describes the intersite exciton hopping
\begin{eqnarray}
H_{ex} = \frac{J_{ex}}{2}\sum\limits_{\left\langle i,j \right\rangle, \sigma} \left[ X^{\sigma s}_i X^{s\sigma}_j + X^{s\sigma}_i X^{\sigma s}_j \right. \nonumber \\
\left. - \left(-1\right)^{\left|\sigma\right|} \left(X^{\sigma s}_i X^{\bar{\sigma}s}_j + X^{s \sigma}_i X^{s \bar{\sigma}}_j    \right) \right],
\label{eq:14}
\end{eqnarray}
where the exciton hopping parameter is 
\begin{equation}
J_{ex}=4\left(t^2_{12}-t_{11}t_{22}\right)/E_g.
\label{eq:15}
\end{equation}
One can note that due to the orthogonality of the HS and LS terms they do not mix locally, but the exciton hopping mix them non locally. The first line in Eq.~\ref{eq:14} describes the intersite single particle exciton hopping, while the second line corresponds to the creation and annihilation of the biexciton pair. We can compare the exciton hopping parameter $J_{ex}$ with similar terms in the effective low-energy models in the literature. In the paper~\onlinecite{Kunes14} the biexciton excitation is possible only due to the interband cross-hopping matrix element $t_{12}$.In the paper~\onlinecite{Nasu} the cross-hopping is not considered, nevertheless the biexciton hopping is possible due to the product $t_1t_2$. As we can see from Eq.~\ref{eq:15}, we have both contributions.

Let us compare two nonlocal parameters of the effective Hamiltonian (\ref{eq:10}), the values of the exchange $J$ (\ref{eq:12}) and exciton hopping $J_{ex}$ (\ref{eq:15}). We consider four different sets of the electron hopping parameters:\\*
A) in the limit $\Delta = \infty$, $t_{12}=t_{22}=0$, we get $J=4t^2_{11}/E_g$ and $J_{ex}=0$ as in the single-band Hubbard model \cite{Anderson}, \\*
B) symmetrical hopping parameters $t_{11}=t_{22}=t_{12}=t$, then the exchange value $J=16t^2/E_g$ is proportional to the superexchange parameter from the Hubbard model, while the exciton hopping $J_{ex}=0$,\\*
C) $t_{12}=0$, then $J=4\left(t_{11}^2+t_{22}^2\right)/E_g$ and $J_{ex}=-4t_{11}t_{22}/E_g$, they have opposite signs,\\*
D) $t^2_{12} \gg t_{11}t_{22}$, then $J=8t_{12}^2/E_g$ and $J_{ex}=4t_{12}^2/E_g$, they are of the same order in magnitude.\\*
These examples and the general expression for the superexchange parameter $J$ demonstrate that antiferromagnetic type of superexchange takes place in our model for all electron hopping parameters, while the hopping of excitons may be positive, negative, and zero.

In the rest of the paper the unimportant term $n_i n_j=4$ for our homopolar case will be omitted from the Hamiltonian. Due to qualitative aim of our paper we will study the effects of the non-Heisenberg contributions and short-order fluctuations given by the spin part (\ref{eq:11}) of the effective Hamiltonian (\ref{eq:10}) with antiferromagnetic exchange parameter, neglecting the exciton dispersion given by the hopping term (\ref{eq:14}). We will restrict ourselves by the symmetrical set B of the hopping parameters, so the exciton hopping parameter~\ref{eq:15} will be zero. Nevertheless, basic exciton processes are still taken into account due to LS term $-\varepsilon_s \sum\limits_{i} X^{ss}_i$ in the Hamiltonian (\ref{eq:11}), which introduces some new non-Heisenberg model effects. Let us illustrate this statement using a simple example. Within MF approximation the Hamiltonian is given by 
\begin{equation}
H_{MF}=\sum\limits_{m=-1}^{1}E_m\sigma X^{mm} - \varepsilon_s X^{ss},
\label{eq:16}
\end{equation}
where $m=-1,0,1$ are triplet states, $z$ is the number of nearest neighbors, $E_m=Jz\sigma m$, so the 3 triplet energy levels $E_m$ are $Jz\sigma$, 0, $-Jz\sigma$, $\sigma$ is the positive sublattice magnetization. Thus, the MF magnetization is
\begin{equation}
\sigma = \frac{\exp(\beta J z \sigma) - \exp(-\beta J z \sigma)}{\sum\limits_{m=-1}^{1}{\exp(-\beta J z \sigma m)} + \exp(\beta \varepsilon_s)},
\label{eq:17}
\end{equation}
which deviates from the Brillouin function due to the LS term. From the other hand, let us consider the exciton Green functions
\begin{equation}
G^m_{ij}=\left\langle \left\langle X_i^{sm}|X_j^{ms}\right\rangle\right\rangle,
\label{eq:18}
\end{equation}
which describe three types of excitons. After writing down the equations of motion and decoupling them using Tyablikov approximation, we have obtained
\begin{equation}
G_m\left(E+i\delta\right)=\left(E-\varepsilon_s+E_m+i\delta\right)^{-1}.
\label{eq:19}
\end{equation}
Thus, the three excitons with spin projection $m=+1,0,-1$ will have the energies $E_{ex}\left(m\right)=E_m-E_s$. This way, at finite temperature the occupation numbers of our HS sublevels can be found from the equation
\begin{equation}
n_m=\left(n_s - n_m\right)f_B\left(E_{ex}\left(m\right)\right),
\label{eq:20}
\end{equation}
where $f_B\left(E\right)$ is the Bose-Einstein distribution function. Together with the completeness condition~\ref{eq:13} we have the full set of MF equations exactly the same as we obtain from Eq.~\ref{eq:17}. This way, we see that in the simplest approximation the exciton process are present in the system and give consistent values for the occupation numbers. Below, instead of MF we will use its cluster generalization, in which all possible positions of singlets within the cluster are taken into account.

\section{\label{sec:3}Cluster mean field theory}

Due to the LS term, the problem given by the Hamiltonian (\ref{eq:10}) cannot be straightforwardly treated by the approaches that work well for the Heisenberg model, like Tyablikov approximation \cite{Bogolyubov, Fu-Cho, Valkov82, Du}, or more sophisticated Green’s function approaches \cite{Kondo, Plakida, Junger04, Junger09}. The simplest approach is to use a MF theory given by the Eq.~\ref{eq:17}. However, the Heisenberg term contains spin fluctuations, which are neglected within the standard MF consideration. To go beyond MF we use its cluster generalization, the self-consistent CMF, which has been applied to various quantum spin models \cite{Valkov06, Valkov07, Brzezicki11, Albuquerque, Brzezicki12, Ren, Gotfryd, Ray, Morita, Koga, Singhania}. We believe that CMF method is suitable for a qualitative study of the toy model we consider at a wide range of temperatures and pressure and it is better anyhow than the single site MF. The approach captures short-range effects, which will be discussed in the next section, and allows treating HS and LS terms equally within a cluster. We note that at high temperature close to second-order phase transition the approach can be considered as only qualitative since it does not capture long-range fluctuations. At zero temperature, as will be presented below, CMF provides results which fall into reasonable agreement with more rigorous approaches.

Within the CMF approach the lattice is covered by translations of a cluster to treat the intracluster interactions by exact diagonalization, whereas the interactions between spins $f$ and $f'$ belonging to different clusters are approximated within MF as ${\bf S}_f{\bf S}_f' \approx S_f^z \left\langle S_f'^z\right\rangle + \left\langle S_f^z\right\rangle S_f'^z- \left\langle S_f^z\right\rangle \left\langle S_f'^z\right\rangle$. Thus, after applying the translational invariance the problem reduces to a single cluster in a MF determined by parameters $\left\langle S_i^z\right\rangle$, which are determined self-consistently by iterative diagonalizations ($i$ runs over boundary sites of a cluster). In our calculations we suppose the mean-fields to be in Neel antiferromagnetic ordering, since there are no competing exchange parameters, but there is a competition between the exchange and the spin gap $\varepsilon_s$, which may be rescaled to pressure. In the main part of the paper we take $J$ as an energy unit and explore the $\varepsilon_s-T$ phase diagram, where $T$ is temperature. For each value of $\varepsilon_s$ and $T$ we compare the free energies of the system in magnetic and non-magnetic phases to decide, which of them is realized. A tolerance factor for convergence of $\left\langle S_i^z\right\rangle$ was set $10^{-5}$. We use full diagonalization at finite temperatures and Lanczos at $T=0$. Since we are dealing with basis consisting of three HS and one LS states, computationaly reasonable sizes of a cluster are $N_c\lesssim10$, where $N_c$ is number of sites, in the former case and $N_c\lesssim20$ in the latter. So, we mostly use a $2\times2$ cluster to illustrate the main physics, but also compare the results using $3\times2$, $4\times2$, and $2\times2\times2$ clusters to study the finite-size effects of our calculations at finite temperature and clusters $4\times3$ and $4\times4$ at zero temperature.

\section{\label{sec:4}Non-Heisenberg behavior and short order effects in the vicinity of spin crossover}

In the main part of this chapter we will discuss the results of our CMF calculations with the spin Hamiltonian (\ref{eq:11}) in the most interesting regime  $\varepsilon_s\sim J$. To compare staggered magnetization obtained with different clusters we will consider the magnetization $m$ on a bulk site, which we define as located as close as possible to the center of a cluster. As known, Fe-based SCO compounds in ambient conditions are 3D magnets. In our cluster calculations it is more numerically practical to consider 2D case, since in 3D only $2\times2\times2$ cluster is available. We can use small $2\times2$ cluster for the main results as well as compare $2\times2$ CMF with larger clusters. Although in 2D the Mermin-Wagner theorem prohibits an ordered state for the spherically symmetric Hamiltonian (\ref{eq:11}), in the case of MF-based approach the results for 2D and 3D are qualitatively identical.

\begin{figure}
\includegraphics{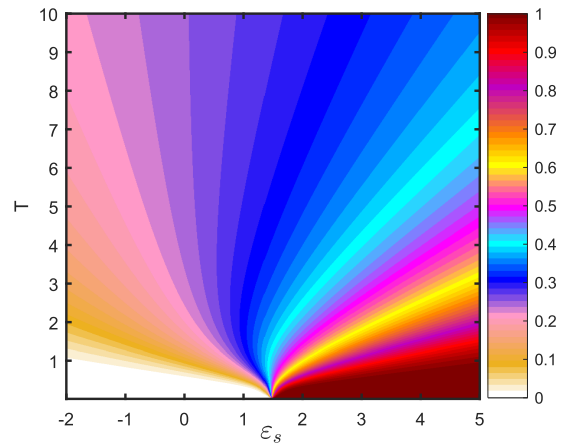}
\caption{\label{fig:2}The map of the LS occupation number obtained with $2\times2$ cluster exact diagonalization.}
\end{figure}

An important quantity characterizing SCO is a HS (LS) concentration. It is accessible in experiments on X-ray emission \cite{Lin} and $\text{M}\ddot{\text{o}}\text{ssbauer}$ spectroscopy \cite{Lyubutin12}. We show in Fig.~\ref{fig:2} the LS concentration $n_{LS}$ dependence on spin gap and temperature obtained by $2\times2$ exact diagonalization. It is qualitatively similar to the obtained experimentaly in Ref.~\cite{Lin} and calculated within MF approaches \cite{Sturhahn, Lyubutin12} and first-principle studies \cite{Tsuchiya}. SCO takes place at $\varepsilon_s = 1.5$ instead of $\varepsilon = 0$ since intracluster exchange interaction stabilizes the HS state and larger crystal field (pressure) is required to reach SCO. Another effect of correlations is the curvature of the isolines of $n_{LS}$ at low temperatures as shown by colors in Fig.~\ref{fig:2}. If to neglect the exchange correlations and take the value $J=0$, all lines of the constant value for LS/HS concentrations will be the straight lines going from the SCO critical point $\varepsilon_s=0$ \cite{Sturhahn, Ovchinnikov11}.

\begin{figure}
\includegraphics{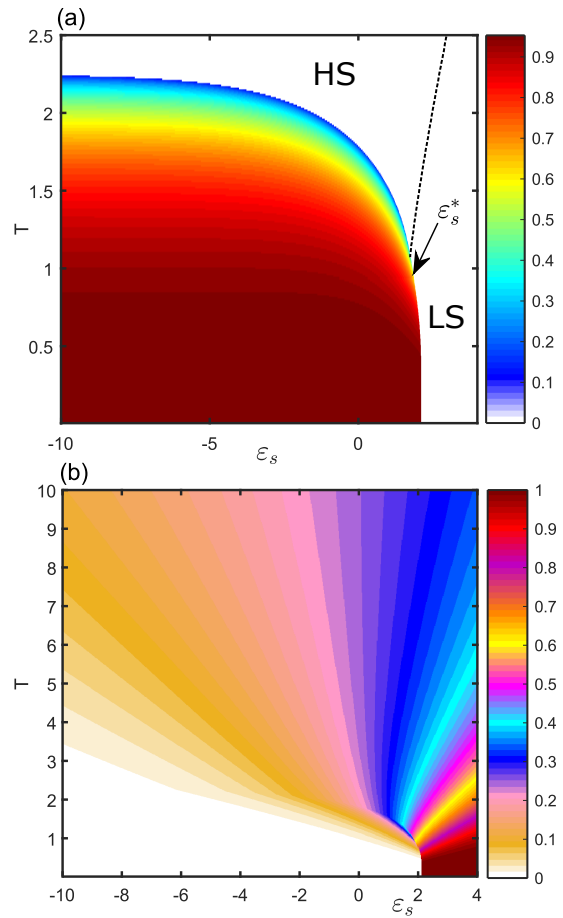}
\caption{\label{fig:3}(a) Average staggered magnetization $m$ and (b) LS occupation number obtained with $2\times2$ CMF. The arrow shows the position of a tricritical point. The dashed line is the $n_{LS} = 0.5$ isoline.}
\end{figure}

As shown in Fig.\ref{fig:3}(a), at $\varepsilon_s\sim-10$ almost Heisenberg behavior of magnetization with temperature is observed, because the system is in the HS state. Thus, a second-order transition from magnetic to nonmagnetic state is realized with heating. From Fig.\ref{fig:3}(b) one can see that for $\varepsilon_s\sim-10$ the population of the LS is zero at low temperature, that provides the conventional Heisenberg model behavior. The nonmagnetic HS phase is the paramagnetic one. With increasing $\varepsilon_s$ thermal fluctuations enhance LS population, so the second-order transition Neel temperature decreases. At $\varepsilon_s=0$ the magnetic transition with heating is still the second order, but the paramagnetic moment is reduced by approximately $20\%$ of the LS states. At $\varepsilon_s = \varepsilon_s^*\approx 1.87$ there is a tricritical point. Increasing $\varepsilon_s$ further leads to a first-order phase transition to nonmagnetic state caused by the change of the ground state from HS to LS, as seen from Fig.~\ref{fig:3}(b). The maximal value of magnetization in Fig.~\ref{fig:3}(a) is $m=0.9528$, instead of $m=1$. This is the manifestation of quantum shortening of spin, which is taken into account partially within CMF by calculating spin-fluctuation terms within a cluster. The non magnetic phase of Fig.\ref{fig:3}(a) can be qualitatively viewed as HS to the left of the $n_{LS}=0.5$ dashed line, which comes out close to the tricritical point, and LS to the right.

\begin{figure}
\includegraphics{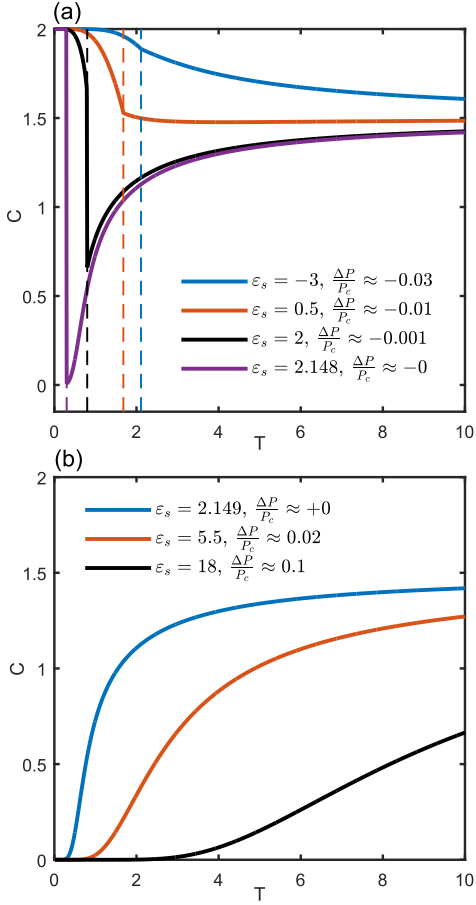}
\caption{\label{fig:4}Temperature dependence of the Curie constant defined by Eq.\ref{eq:16} for different values of spin gap (pressure) obtained with $2\times2$ CMF (a) below, (b) above the critical pressure. The dashed lines indicate the values of the Neel temperature for the data of the same color, $\mu^2$ of Eq.~\ref{eq:16} is set equal to one.}
\end{figure}

The distribution of LS density in Fig.\ref{fig:3}(b) is related to the Curie constant in paramagnetic susceptibility
\begin{equation}
C = \mu^2\left(1-n_{LS}\right)S\left(S+1\right),
\label{eq:21}
\end{equation}
where $\mu = \frac{\mu_B^2}{3k_B}$. The temperature dependence of $C$ is shown in Fig.~\ref{fig:4} for different values of the spin gap. Equation (\ref{eq:16}) makes sense for the paramagnetic phase above the Neel temperature indicated in Fig.~\ref{fig:4}(a) by dashed lines. Using parameters extracted from the anvil-cell experiments on ferropericlase \cite{Lyubutin12, Lyubutin13} we can estimate the corresponding values of pressure $P$ by assuming that the spin gap defines pressure as $\varepsilon_s-\varepsilon_s^c= \alpha_\Delta(P-P_c)$, where $\alpha_\Delta=7.8\text{meV}/\text{GPa}$, the critical pressure $P_c$ is $55\text{GPa}$ and taking into account the pressure dependence of the exchange integral is $J\left(P\right)=J_0\left(1+\frac{2\alpha_t}{t}P\right)$, where $J_0$ is taken to be $18\text{K}$ and $\frac{2\alpha_t}{t}=0.0122 1/\text{GPa}$. This way, for each value of $\varepsilon_s$ we show corresponding pressure values $\frac{\Delta P}{P_c} = \frac{\left(P-P_c\right)}{P_c}$. Note that within this set of parameters the exchange integral value is chosen to reproduce the real compound's Neel temperature and the critical pressure is aligned with our critical value of the spin gap for a more convenient qualitative discussion of our results in a context of experimental data as discussed below. Few percent below the critical pressure there is simply a drop of an effective magnetic moment with temperature. Around percent below $P_c$ an effective magnetic moment is almost temperature independent. Very close to critical pressure the LS component at the Neel temperature is already significant and thermal fluctuations lead mainly to increase of the HS component. Above the critical pressure, as shown in Fig.~\ref{fig:4}(b), increasing pressure leads to slowdown in temperature growth of an effective magnetic moment.

\begin{figure}
\includegraphics{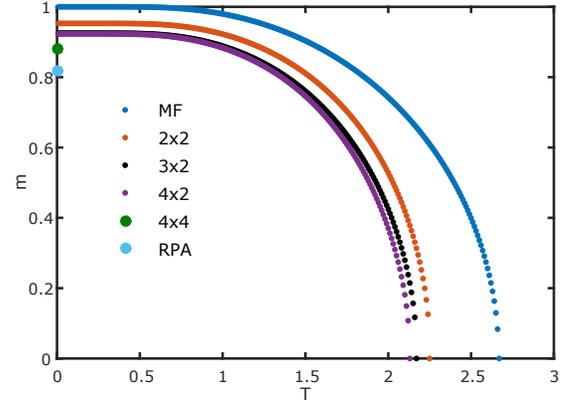}
\caption{\label{fig:5} Bulk site's magnetization calculated in the Heisenberg limit in 2D within MF, CMF with different rectangular clusters, and RPA.}
\end{figure}

\begin{figure}
\includegraphics{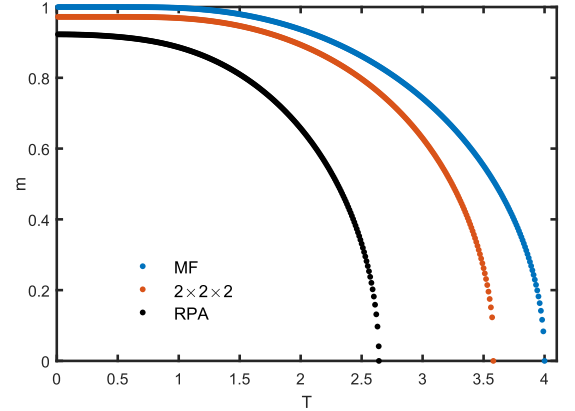}
\caption{\label{fig:6}The same as in Fig.~\ref{fig:5} in 3D within MF, $2\times2\times2$ CMF, and RPA.}
\end{figure}

\begin{figure}
\includegraphics{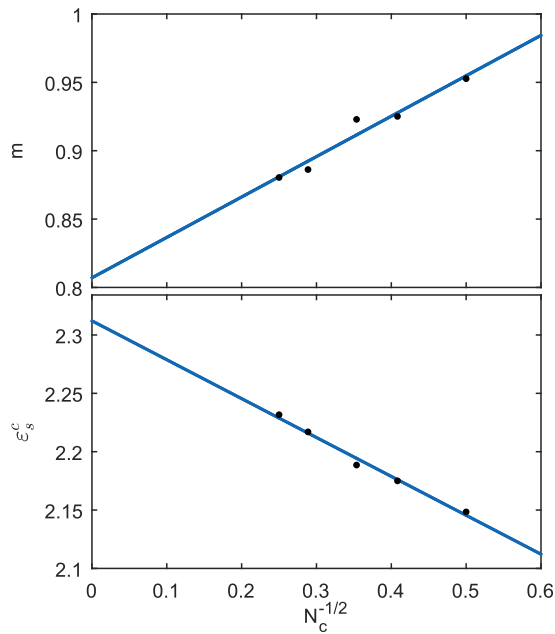}
\caption{\label{fig:7}Extrapolation to the infinite system limit for the bulk magnetization $m(N_c)$ and the critical value of the spin gap $\varepsilon^c_s(N_c)$ in the Heisenberg limit.}
\end{figure}

To explore finite-size effects of our CMF calculations we now turn to comparison of magnetization obtained within different clusters and within the Tyablikov approximation (or RPA) in the Heisenberg limit. Within the Heisenberg model RPA is known to provide results in a decent agreement with numerically exact quantum Monte Carlo \cite{Junger09, Yasuda05}. From Fig.~\ref{fig:5} it is seen that inclusion of nearest correlations leads to an appearance of zero fluctuations in $m$ and a substantial decrease in Neel temperature when comparing MF with $2\times2$ CMF. At zero temperature the bulk magnetization seems to gradually approach the RPA value 0.8168, for example for $4 \times 3 $ (not shown) and $4 \times 4$ clusters we obtain $m=0.886$ and $m=0.88$.  In 2D the Neel temperature is zero in RPA, since it satisfies to the Mermin-Wagner theorem, unlike (C)MF, where the symmetry of the cluster’s (site’s) Hamiltonian is lowered artificially. Analogous comparison in 3D is shown in Fig.~\ref{fig:6}: the Neel temperature is approximately 1.5 times higher within MF that within RPA and 1.33 times higher with $2\times 2\times 2$ CMF. This way, in terms of staggered magnetization’s and Neel temperature’s values we obtain intermediate results between RPA and MF. In our CMF calculations in the 2D case the bulk site magnetization $m(N_c)$ as a function of the number of sites  turned to be proportional to $\sqrt{N_c}$. Least square extrapolation gave the result $m_{\infty}\approx0.81$, which is similar to the RPA value (see Fig.~\ref{fig:7}).

\begin{table}
\caption{\label{tab:1}
Tricritical $\varepsilon^*_s$ and critical $\varepsilon^c_s$ values of the spin gap for different clusters within CMF.}
\begin{ruledtabular}
\begin{tabular}{c|cccccc}
&
\multicolumn{1}{c}{\textrm{$MF$}}&
\multicolumn{1}{c}{\textrm{$2\times2$}}&
\multicolumn{1}{c}{\textrm{$3\times2$}}&
\multicolumn{1}{c}{\textrm{$4\times2$}}&
\multicolumn{1}{c}{\textrm{$4\times3$}}&
\multicolumn{1}{c}{\textrm{$4\times4$}}\\
\hline
$\varepsilon^*_s $ & $\approx 1.59$ & $\approx 1.87$ & $\approx 1.93$ & $\approx 1.98$ & ---\footnote{Have not been calculated for this cluster.} & ---\\
$\varepsilon^c_s $ & 2 & 2.148 & 2.175 & 2.189 & 2.217 & 2.232\\
\end{tabular}
\end{ruledtabular}
\end{table}

\begin{figure}
\includegraphics{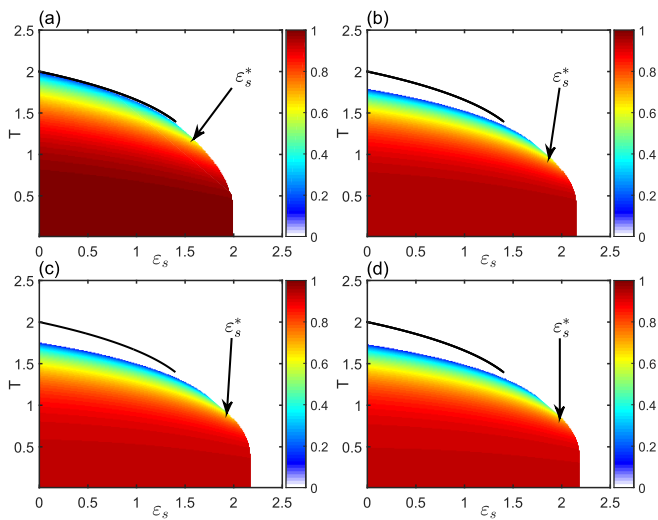}
\caption{\label{fig:8} Bulk site's magnetization obtained within (a) MF, (b) $2\times 2$, (c) $3\times 2$, and (d) $4\times 2$ CMF. The black line shows MF second-order transition line. Arrows show the position of a tricritical point.}
\end{figure}

Next, we compare average staggered magnetization obtained with different clusters and MF at different values of spin gap in Fig.~\ref{fig:8}. Phase diagrams obtained within different clusters are very similar. Besides the decrease in Neel temperature there is an increase in tricritical value of a spin gap $\varepsilon^*_s$ and the critical value $\varepsilon^c_s$, at which the first-order phase transition occurs, as it is shown in Table~\ref{tab:1}. The increase in $\varepsilon^c_s$ with cluster's size is related to the lowering of the cluster’s ground state energy in magnetic phase with increasing size, because the main competition is between states with 0 and $N_c$ singlets per cluster. Similarly to the case of magnetization, we observe $1/\sqrt{N_c}$ behavior of $\varepsilon^c_s(N_c)$ or the ground-state energy $E_0$ with opposite sign in the Heisenberg limit (see Fig.~\ref{fig:7}). By least squares extrapolation for $E_0(N_c)$ we found $E_0(\infty) \approx -2.31$, which is similar to the value $E_0(\infty) \approx -2.33$ from the quantum Monte Carlo \cite{Harada} and density matrix renormalization group \cite{Ramos} studies. The size dependence of $\varepsilon^*_s$ and $\varepsilon^c_s$ shows the most crucial change when going from MF to 4-site CMF with predictable behavior when increasing the system size. Thus, the part of the phase diagram obtained at finite temperature close to the first order transition with small clusters from 4 to 8 sites can be considered as semi-quantitative.

\begin{figure}
\includegraphics{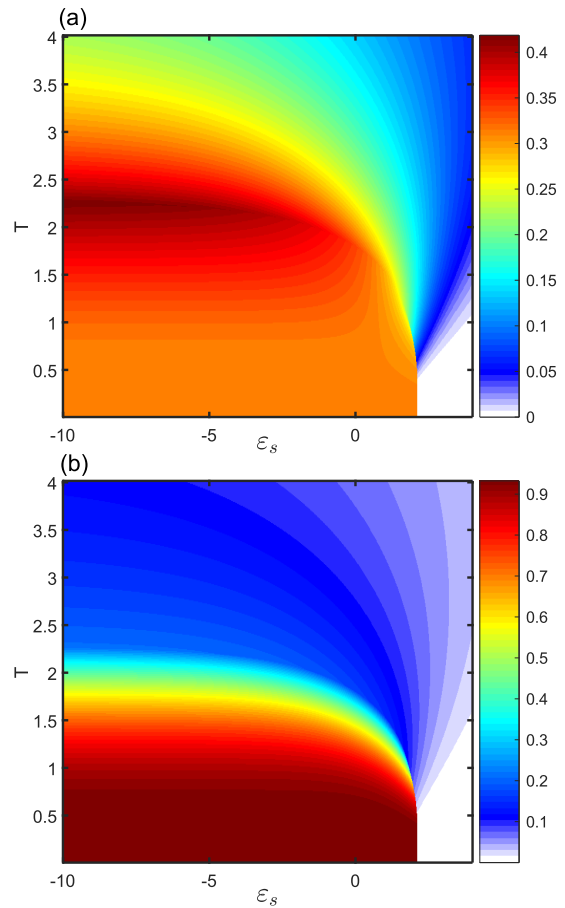}
\caption{\label{fig:9}(a) Transverse $C_{\bot}$ and (b) longitudinal $C_{\parallel}$ nearest-neighbor spin correlations, obtained within $2\times 2$ CMF.}
\end{figure}

Although within standard MF approach qualitatively correct magnetic phase diagram is obtained, it provides no information about short-range correlations in the system. In Fig.~\ref{fig:9} we show transverse antiferromagnetic nearest-neighbor spin correlations $C_{\bot}=-\left\langle \left(S^+_0S^-_1 + S^-_0S^+_1\right)\right\rangle$ and longitudinal ones $C_{\parallel}=-\left\langle S^z_0S^z_1\right\rangle$.  At $\varepsilon_s<\varepsilon^c_s$ the longitudinal correlations are always decreasing with temperature, but transverse ones are increasing with temperature at low values of spin gap, reaching maximum at Neel points and lowering in a paramagnetic phase. A non-Heisenberg effect is that at $\varepsilon_s>\varepsilon^c_s$ the spin correlations show a reentrant behavior. At low temperature they are zero, then increasing with heating due to thermal excitement of triplet states. When temperature is increased further, the correlations lower again.

\begin{figure}
\includegraphics{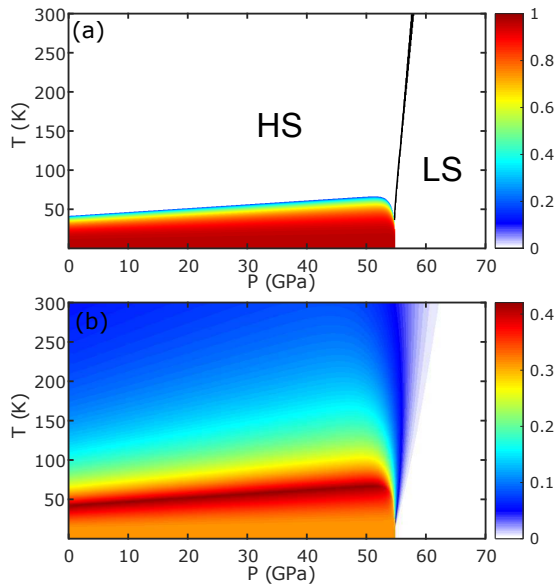}
\caption{\label{fig:10} (a) Average sublattice magnetization $m$ calculated for ferropericlase parameters from Ref.~\cite{Lyubutin12} by $2\times 2$ CMF. The black line is the $n_{LS} = 0.5$ isoline. (b) Transverse spin correlations for the same set of parameters.}
\end{figure}

Finally, we use parameters from the anvil-cell experiments on ferropericlase (Mg,Fe)O ~\cite{Lyubutin12, Lyubutin13} used above to model its magnetization dependence on pressure and temperature. The exchange parameter value and its linear pressure dependence at low pressure in the HS state were obtained by fitting the experimental data from the paper~\onlinecite{Lyubutin12}. The magnetization’s phase diagram is presented in Fig.~\ref{fig:10}(a). Heisenberg behavior is realized in a broad range of pressure, where the Neel temperature scales linearly with pressure and reaches its maximum. At $P\approx P_c$ the Neel temperature drops discontinuously to zero due to a phase transition of the first order. Deviation from Heisenberg behavior is realized at $P \gtrsim 51 \text{GPa}$ at $T=0$ and at $P \gtrsim 45 \text{GPa}$ at room temperatures, as it is seen from spin correlations in Fir.~\ref{fig:10}(b). The non magnetic phase can be qualitatively identified as HS to the left of the black line, which denotes $50\%$ of maximal effective magnetic moment, and LS to the right. Our phase diagram is consistent with experimental data and model calculations of Refs.~\cite{Lyubutin12, Lyubutin13}. This shows that the microscopic Hamiltonian we have studied is capable of capturing the main physics of spin crossover in ferropericlase.

\section{\label{sec:5}Discussion}

To sum up, in order to study non-Heisenberg effects due to SCO we have derived an effective Hamiltonian for the two-orbital Kanamori model. The parameters of the effective Hamiltonian have been written down analytically. It contains HS and LS states, and interatomic exchange interaction, as well as the exciton hopping and the biexciton creation and annihilation processes. As it can be seen within simple MF, due to the presence of LS states the MF magnetization within this model is not described by the Brillouin function. The effective Hamiltonian has been studied within CMF approximation. As we have shown by comparing our results between different cluster sizes and to other methods in the special case, our results are of qualitative character at high temperatures, but we expect them to be semi-quantitative within an interesting region close to first-order transition. We have obtained a magnetic $\varepsilon_s-T$ phase diagram of the model with antiferromagnetic and paramagnetic phases. At very low spin gap  values $\varepsilon_s$  the magnetization’s temperature dependence is almost Heisenberg-like. Increasing $\varepsilon_s$ leads to reduction of the Neel temperature and paramagnetic moments (or the Curie constant in the paramagnetic susceptibility) due to thermal population of LS states. Up to a tricritical point $\varepsilon^*_s$ the phase transition line is second-order one and from $\varepsilon^*_s$ to a critical value of quantum phase transition $\varepsilon^c_s$ it is first-order. Few percent below $\varepsilon_s$ there occurs a drastic change in the temperature dependence of the Curie constant in paramagnetic susceptibility. At $\varepsilon_s>\varepsilon^c_s$ the magnetic moment and the Curie constant are zero at zero temperature and they increase with heating because of growing population of HS states. From quantitative point of view we expect our results for the magnetic phase diagram to be between simple MF (closer to MF) and RPA, which has not been rigorously developed yet in the case when LS states must be taken into account. However, we have shown that the results of CMF calculations shall approach correct values with further increase in cluster's size, thus showing predictable behavior. Using cluster approach has allowed us to predict another non-Heisenberg effect, which is a reentrant behavior of the temperature dependence of spin correlation functions at $\varepsilon_s>\varepsilon^c_s$. For the $P-T$ magnetic phase diagram that we have obtained for ferropericlase the non-Heisenberg behavior is realized at $P \gtrsim 51 \text{GPa}$ at $T=0$ and $P \gtrsim 45 \text{GPa}$ at $T\approx300K$, which is a realistic pressure and temperature interval for a more detailed experimental investigation of this compound and for observing the non-Heisenberg effects.

\begin{acknowledgments}
The authors thank the Russian Scientific Foundation for the financial support under the grant 18-12-00022.
\end{acknowledgments}

\bibliography{Paper}

\end{document}